\documentstyle[prl,aps,multicol]{revtex}
\begin{document} 
\draft 
\title{Photoemission spectra of the  $t-J$ model in one and two dimensions:
similarities and differences}

\author{R. Eder$^1$ and  Y. Ohta$^2$}
\address{$^1$Department of Applied and Solid State Physics,
University of Groningen, 9747 AG Groningen, The Netherlands\\
$^2$Department of Physics, Chiba University, Inage-ku, Chiba 263, 
Japan} 
\date{\today}
\maketitle

\begin{abstract} 
We present an exact diagonalization study of the single particle spectral 
function in the 1D and 2D $t$$-$$J$ model. By studying the scaling 
properties with $J$ and $t$ we find a simple building pattern 
in 1D and show that every spectral feature can be uniquely assigned by a 
spinon and holon momentum. We find two types of low energy excitations: 
a band with energy scale $J$ and high spectral weight disperses upwards 
in the interior part of the Brillouin zone and reaches $E_F$ at $k_F$, 
and a band with energy scale $t$ and low spectral weight disperses downwards 
in the outer part of the zone, touching $E_F$ at $3k_F$. 
An analogous analysis of the 2D case at half filling shows that the
$t$-band exists also in this case, but is diffuse and never
reaches the Fermi energy. For the doped case in 2D the picture is
more reminiscent of 1D, in particular the `main-band'
with a dispersion $\propto J$ and the `shadow band' with energy scale $t$
can be identified also in this case. This leads us to propose 
that the shadow bands discovered by Aebi {\em et al.} in Bi2212 are the 2D 
analogue of the $3k_F$ singularity in 1D systems and unrelated to
antiferromagnetic spin fluctuations.
\end{abstract} 
\begin{multicols}{2}

\section{Introduction}
Recent developments in photoelectron spectroscopy have challenged the
apparent simple truth that the Fermi surface of cuprate
superconductors is simply the one corresponding to LDA
band structures with the only effect of the closeness to the Mott-Hubbard 
insulator being a moderate correlation narrowing of the band width.
The discovery of the `shadow bands'\cite{Aebi,LaRosa}, the 
temperature dependent pseudogap in the
underdoped state\cite{Loeser} and the substantial doping dependence of the
quasiparticle band structure\cite{Marshall} leave little doubt that
a simple single-particle description is quite fundamentally
inadequate for these materials. Moreover, photoemission experiments on
one-dimensional (1D) copper oxides\cite{Kim} have shown very clear 
signatures of spin charge separation. The equally clear nonobservation
of these signatures in the cuprate
superconductors at any doping level advises against another
apparent simple truth, namely that the Fermi surface seen in the cuprates
is simply that of the `spinons' in a 2D version of the
Tomonaga-Luttinger liquid (TLL) realized in 1D.
Motivated by these developments, we have performed a detailed
exact diagonalization study of the electron removal spectrum in the
1D and 2D $t$$-$$J$ model. This model reads
\[
H = -t \sum_{\langle i,j \rangle,\sigma}
(\hat{c}_{i,\sigma}^\dagger \hat{c}_{j,\sigma} + {\rm H.c.}) + J
\sum_{\langle i,j \rangle} \big( \vec{S}_i \cdot \vec{S}_j 
-{1\over 4} n_i n_j \big).
\]
There by the `constrained' Fermion operators are written as
$\hat{c}_{i,\sigma} = c_{i,\sigma} n_{i,\bar{\sigma}}$ and
$\vec{S}_i$ denotes the spin operator on site $i$.
The summation $\langle i, j \rangle$ extends over all pairs
of nearest neighbors in a 1D or 2D square lattice.\\
The electron removal spectrum is defined as
\[
A(\vec{k},\omega) = \frac{1}{\pi} \Im
\langle \Psi_0 | \hat{c}_{\vec{k},\sigma}^\dagger
\frac{1}{ \omega - (E_0 - H) - i0^+} \hat{c}_{\vec{k},\sigma}
|\Psi_0\rangle,
\]
where $E_0$ and $|\Psi_0\rangle$ denote the ground state
energy and wave function. For small finite clusters,
this function can be evaluated numerically by means of the
Lanczos algoritm\cite{Dagoreview}. \\
In 1D the $t$$-$$J$ model is solvable by Bethe ansatz in the case
$J$$=$$2t$\cite{BaresBlatter}, but even for this limit the
complexity of the Bethe ansatz equations precludes an evaluation of
dynamical correlation functions.
For the closely related Hubbard model in the
limit $J/t$$\rightarrow$$0$ the Bethe-ansatz
equations simplify\cite{OgataShiba}, and an actual calculation of the
spectral function becomes possible\cite{SorellaParola,Penc}. 
In all other cases Lanczos diagonalization is the only
way to obtain accurate results for $A(\vec{k},\omega)$\cite{Favand}.\\
In order to analyze our numerical results, we first want to develop an
intuitive picture of the scaling properties of the elementary excitations
in 1D, which will turn out to be useful also in 2D.
It has been shown by Ogata and Shiba\cite{OgataShiba} that for  
$J/t$$\rightarrow$$0$ the wave functions can be constructed
as products of a spinless Fermion wave function, which depends only
on the positions of the holes, and a spin wave function, 
which depends only on the sequence of spins.
A naive explanation for this remarkable property
is the `decay' of a hole created in a N\'eel ordered
spin background into an uncharged spin-like domain wall,
and a charged spinless domain wall.
Then, since it is the kinetic energy $\sim$$t$ which propagates
the charge-like domain walls, whereas the exchange energy $\sim$$J$
moves the spin-like domain walls, one may expect that the two
types of domain walls have different energy scales.
Namely the excitations of the charge part of the wave function
(i.e., the `holons') have $t$ as their energy scale, whereas those of the
spin part (i.e., the `spinons') have $J$ as their energy scale.
Scanning the low energy excitation spectrum of 1D $t$$-$$J$ rings
then shows that indeed most of the excited states
have excitation energies of the form $a\cdot t + b\cdot J$\cite{TK},
which  indicates the presence of two different
elementary excitations with different energy scales.\\
Surprisingly enough the low energy spectrum of the 2D model
shows the same scaling behavior of the excitation energies
as in 1D\cite{TK},
which seems to indicate the existence of two types of
spin and charge excitations if very different nature
also in this case. Other cluster results
indicate, however, that these two types of excitations
do not exist as `free particles':
the dynamical density correlation function,
which corresponds to the `particle-hole excitations' of holons
and shows sharp low energy peaks in 1D\cite{BaresBlatter}
is essentially incoherent in 2D and has practically no sharp low energy
excitations\cite{EderOhtaMaekawa}. The optical
conductivity in 2D shows an incoherent high energy part
with energy scale $J$\cite{EderWrobelOhta} - which is completely unexpected
for the correlation function of the current operator which acts only 
on the charge degrees of freedom. There is moreover rather clear
numerical evidence\cite{DagottoSchrieffer,EderOhta,RieraDagotto} 
that the hole-like low 
energy excitations can be described to very good approximation as 
spin $1/2$ `spin bags'\cite{Schrieffer}
 - i.e., holes dressed heavily by a local cloud of spin excitations.\\
To obtain further information about similarities and
differences between 1D and 2D, also in comparison to
the spectroscopic results, we have performed a systematic
comparison of the electron removal spectra in both cases.
As will become apparent, there are some similarities, but also
clear differences. We suggest that the main difference between
1D and 2D is a strong attractive interaction between `spinon' and
`holon' in 2D, which leads to a band of bound states being pulled out
of the continuum of free spinon and holon states. This band
of bound states - which are nothing but simple spin $1/2$ Fermions
corresponding to the doped holes - then sets the stage for the
low energy physics of the system, i.e., true spin-charge separation
as in 1D never occurs.

\section{One dimension, half filling}

We begin with a discussion of the 1D model at half-filling.
Figure \ref{fig1} shows the
electron removal spectra for the $12$-site ring.
Let us first consider 
the left panel, where energies are
measured in units of $J$. Then, one can distinguish
different types of states according to their scaling behavior with
$t$: there is one `band' of peaks (connected by
the thin full line) whose energies relative
to the single-hole ground state at $k$$=$$\pi/2$ remains practically
unchanged under a variation of $t$, i.e., these states have $J$
as their energy scale. As a remarkable fact,
this `band' abruptly disappears half-way in the Brillouin zone,
i.e., there are no peaks whose energy scales with $J$
beyond $k$$=$$\pi/2$. This looks like a half-filled free-electron band
with a Fermi level crossing at $\pi/2$, which however is quite remarkable
because inverse photoemission is not possible at half-filling.
Next, in addition to this `$J$-band', there are several
groups of peaks whose excitation energy shows a very systematic progression
with $t$. Indeed, when plotting the same spectra but
measuring energies in units of $t$ (right panel of Fig.~\ref{fig1})
these peaks 
coalesce, i.e., to excellent approximation the energy scale
of these states is $t$. This coexistence of states with different energy
scales can be nicely seen in the `double peak' for $t/J$$=$$2$ and
momentum $2\pi/6$: the peak with lower binding energy
falls into the $J$-band, the one with the higher binding
energy belongs to the $t$-band. The dispersion of the $t$-band
resembles a slightly asymmetric parabola with minimum
near $\pi/2$ for the low excitation energies that we are considering.
The states that fall onto this parabola
correspond to the creation of a spinon with momentum
$k_F$$=$$-\pi/2$, and a holon of momentum $k+\pi/2$. Since
the spinon momentum is fixed, this group of states 
then simply traces out the holon dispersion. On the other hand,
the `$J$-branch' corresponds to the holon momentum being
fixed at the minimum of the holon dispersion,
and thus traces out the spinon dispersion.\\
This building principle for the spectra can be pushed further.
Namely, one might expect that not only $k_F$ but any spinon momentum
may serve as the starting point for
a complete branch of peaks which trace out
the full holon dispersion. That this is indeed the case
is shown in Fig.~\ref{fig2}. There, the entire width of the
spectra is shown and we have chosen the
zero of energy at the excitation energy of
either the topmost `$J$-peak' at $k_F=\pi/2$ (left panel)
or the topmost `$J$-peak' at $\pi/3$ (right panel).
Due to this choice of the zero of energy,
the energy $\propto J$ of the spinon with the respective
momentum drops out. Then,
when measuring energies in units of $t$ different holon bands
`become sharp', i.e., their energy
{\em relative to the respective spinon energy} scales accurately with $t$.
Moreover, these different
groups of peaks to good approximation
all trace out the same simple backfolded nearest neighbor hopping
dispersion, i.e., the dispersion of the holon is simply
$2t \cos(k_x)$. As discussed above, the 
first holon band is shifted by the spinons' Fermi momentum,
$k_F=\pi/2$, so that its dispersion near the band minimum
at $k=\pi$ could be seen in Fig.~\ref{fig1}.
We have also verified that by alligning the spinon peaks at
$\pi/6$ yet another complete holon band can be identified.\\
We can thus infer the following building principle
for the spectral function: the basis for the whole construction is the
`half-filled' spinon band, with dispersion $-0.65 J \cos(k)$; this
is indicated by the thick dashed line in Fig.~\ref{fig3}a. Then,
each $(k,\omega)$-point of this band provides the `basis' for
a complete holon band $2 t \cos(k)$,
which is `hooked on' to the spinon band at its band maximum;
these holon bands are indicated by the thin full lines in Fig.~\ref{fig3}(a).
Comparison with the numerical results (in this
case for the $20$-site ring) in Fig.~\ref{fig3}(b) shows that indeed to
excellent approximation the poles of the single particle spectral
function fall onto these bands. There are some
deviations at high binding energies, which however are most probably
a deficiency of the Lanczos spectra, which are highly accurate only at
low excitation energy. Moreover, the holon bands in Fig.~\ref{fig3}(b) 
have been slightly shifted, i.e., they are `hooked on' to the
spinon band not precisely at their maximum - we have verified that
this shift has oscillating sign for different chain lengths,
so that it probably is a finite size effect.
As an interesting feature, the pole strength
seems to be constant along each of these holon bands,
i.e., the weight is a function only of the spinon momentum
(this seems not to be correct for $k$$=$$0$ and $k$$=$$\pi$; here it should 
be noted that for these momenta the holon band intersects itself,
which leads to a doubling of the peak weight).
In the thermodynamic limit, the density of bands increases,
while simultaneously their spectral weight decreases, resulting
in incoherent continua. Comparing with the exact results of
Sorella and Parola\cite{SorellaParola} for the case $J/t$$\rightarrow$$0$ it 
is obvious that the outermost holon band in our calculation,
originating from the spinon Fermi momentum,
develops into a cusp-like singularity of the spectral weight.
The spinon band itself, whose energy scale is $J$,
turns into a second dispersionless
cusp in this limit, which skims at zero excitation energy
between $k$$=$$0$ and $k$$=$$\pi/2$. Sorella and Parola
found the excitation energy of the dispersive cusp to be
$-2t \pm 2t\sin(k)$, which  corresponds to a
the backfolded and shifted nearest neighbor hopping band,
$-2t + 2t \cos(k-\pi/2)$. \\
Summarizing the data for 1D we see that the entire
electron removal spectrum obeys a very simple
building principle, which moreover holds for all momenta 
and frequencies.
Analyzing the scaling of the different features
with $J$ and $t$ one can  identify `branches' of states
which trace out the dispersion of the true elementary excitations
of the TLL, namely the collective spin and charge excitations.
The dispersions of the spinons and holons are both
consistent with simple nearest neighbor hopping bands,
the spinons moreover have a half-filled Fermi surface.
While these results may not be really new or surprising, we note that
they demonstrate that  exploiting the scaling properties of excitation
energies provides a very useful method to identify
the different `subbands'. In the following, we will make
extensive use of this principle to address the far less understood
problems of 2D and finite doping.

\section{Two dimensions, half filling}

We proceed to the 2D model, and also consider first the case of 
half filling.
The spectra shown below refer to the standard $20$-site cluster,
which is the largest cluster for which the
calculation of the electron removal spectrum is feasible
also in the doped case. The $\vec{k}$-net for this cluster,
which is shown in Fig.~\ref{fig3a}, consists of the group of
momenta which roughly follows the $(1,1)$ direction,
and a second group along $(\pi,0)$$\rightarrow$$(0,\pi)$.
We would like to stress that results for other clusters are
completely consistent with those for the $20$-site cluster.
Then, the left panel of Fig.~\ref{fig4} shows the photoemission spectrum
for this cluster at half-filling; thereby we again focus on
energies within a few $J$ from the top of the
band and measure energies in units of $J$.
When the spectra are aligned at the top of the band, the positions
of the other dominant low energy peaks do not show a strong variation with
$t$. Some peaks do show a slight but systematic drift with $t$,
which however is much weaker than in 1D. A peculiar feature is
the peak at $(0,0)$, whose relative excitation energy
decreases rather than increases with $t$. Inspection shows, however,
that the (very weak) dispersion along the line
$(\pi,0)$$\rightarrow$$(0,\pi)$ (i.e., the lowest three momenta
in Fig.~\ref{fig4}) scales with $t$ to good approximation.
A possible explanation is the fact
that a hole in a 2D system has two distinct
mechanisms for propagation, firstly by `string truncation',
which gives effective hopping integrals $\sim$$J$, and secondly
by hopping along spiral paths\cite{Trugman}, which gives (smaller)
effective hopping integrals $\sim$$t$\cite{EderBecker}.
It can be shown\cite{EderBecker} that the 
dispersion relation for a single hole to good approximation can be
written as
\[
E(\vec{k}) = J \cdot c_1 (cos(k_x) + \cos(k_y))^2
- t \cdot c_2 \cos(k_x)\cos(k_y)
\]
where $c_1\gg c_2 >0$ are numerical constants. The first term,
which originates from the string truncation mechanism,
gives a dispersion which is degenerate
along $(\pi,0)$$\rightarrow$$(0,\pi)$ and this
degeneracy is lifted by the second term which is the contribution from the
spiral paths; this naturally explains the scaling of 
the dispersion along this line with $t$.
Comparing with 1D we note that with the exception of $(\pi,\pi)$
the `$J$-band' is present in the entire Brillouin zone,
i.e., the spinon Fermi surface seen at half-filling
in 1D does not exist. \\
We turn to the right panel of Fig.~\ref{fig4}(b), which shows the entire
width of the spectra, with energies measured in units of $t$.
It is first of all quite obvious that the spectra generally
are more `diffuse' than in 1D, with sharp features existing
only in the immediate
neighborhood of the top of the band (except for one relatively sharp
high energy peak at $(\pi,\pi)$). Next, among the diffuse features
at high energy there are some whose energy accurately
scales with $t$. Although these `peaks' are rather broad, 
so that the assignment of a dispersion is not really meaningful,
their centers of gravity can be roughly fitted by the expression
$-2t \pm 2t\sin(|k_x| + |k_y|)$, which is reminiscent of the
dispersion of the `holon-cusp' found by Sorella and 
Parola\cite{SorellaParola}
in 1D. An important difference as compared to 1D
is the fact that this $t$-band does not seem to
reach the top of the photoemission spectrum - rather it
stays an energy of $\sim$$t$ below the $J$-band, which forms the
first ionization states.
We believe that `in 1D language' the most plausible interpretation
of the data is the formation of bound states of
spinon and holon: assuming a strong attraction between
these two excitations, which may originate, e.g., from the
well-known string mechanism for hole motion in an 
antiferromagnet\cite{Bulaevskii}, one may expect that a band of bound states
is pulled out of the continuum of free spinons and holons.
This band of bound states
corresponds to the $J$-band (which however has a small
contribution $\propto t$ in its dispersion due to the
spiral path mechanism). Such a bound state
of spinon and holon should be a spin-bag--like spin $1/2$ Fermion,
i.e., a hole heavily dressed by spin excitations.
There is strong numerical 
evidence\cite{DagottoSchrieffer,EderOhta,RieraDagotto}, 
that this is indeed the
character of the low energy states in 2D at low doping.
One may expect, however, that such a bound state may not be stable
for all momenta, and we believe that this
is the reason for the absence of a $J$-peak at $(\pi,\pi)$.
In this picture, the 2D analogue of the holon 
is not a coherently propagating excitation, because it is bound to
the much slower spinon by the linearly ascending string potential. This
picture fits nicely with the diffuse character of the
dynamical density correlation 
function in 2D\cite{EderOhtaMaekawa}: this function,
which in a TLL should measure basically the response of the free
holons, in 2D has almost exclusively diffuse high energy `peaks',
with virtually no sharp low energy peaks.
Moreover, the unexpected (in the framework
of spin charge separation) appearance of $J$ as energy scale
in the optical conductivity is also readily understood
in terms of the dipole-excitations of a bound spinon-holon 
pair\cite{EderWrobelOhta}.\\
Summarizing the data for 2D, we see a band of quasiparticle peaks,
which predominantly has $J$ as its energy scale,
and some diffuse high energy `band' with energy scale $t$.
Both, the absence of the `spinon Fermi surface', as well as the lack
of sharp `holon bands' are in clear contrast to the situation in 1D.
The formation of bound states of spinon and holon, resulting in
a split-off band of spin-bag--like spin $1/2$ Fermions 
explains this in a natural way.

\section{One dimension, doped case}

We return to 1D and consider the doped case.
Figure \ref{fig5} shows the spectral function for the
$12$-site ring with $2$ holes. Measuring excitation energies
in units of $J$ (left panel) we can again identify the spinon band.
For $2$ holes in $12$ sites the nominal Fermi momentum is
$k_F$$=$$5\pi/12$ (i.e., half way between $\pi/3$ and $\pi/2$)
and the spinon band extends up to this momentum.
As was the case at half-filling,
some other peaks show a systematic progression of their
excitation energy, and switching the unit of energy to $t$
(right panel) again makes a nearly complete `holon band' visible to
which these peaks belong. The holon band again takes the form
of a backfolded tight-binding band, but this time the
top of the parabola around $k$$=$$\pi/2$ is missing.
The holon band now
seems to touch the Fermi energy at $k_F$ and at $3k_F$$=$$9\pi/12$
(the latter momentum is half way between $4\pi/6$ and $5\pi/6$).
This picture of the spectral function nicely fits with the recent
exact calculation in the limit $J/t$$\rightarrow$$0$
by Penc {\em et al.}\cite{Penc}:
on the photoemission side, this calculation showed
a high intensity `band' which is very similar to the
backfolded tight-binding dispersion of the holon
band. In addition there was a dispersionless low intensity
band at zero excitation energy, which 
corresponds to the spinon band in the limit $J/t$$\rightarrow$$0$.
For both, the exact result in the limit $J/t$$\rightarrow$$0^+$,
and our numerical data for finite $J$, there are thus
two branches of states which reach excitation energy zero:
the `main band' which touches $E_F$ at $k_F$,
and the `shadow band'\cite{Favand}, which reaches $E_F$ at $3k_F$.
The `Fermi level crossings' of these two bands may be thought of
producing the well known (marginal) singularities
in the electron momentum distribution $n(k)$ at $k_F$ and $3k_F$,
found by Ogata and Shiba\cite{OgataShiba}. \\
The numerical spectra demonstrate
a peculiar feature of the TLL, namely a kind of Pauli
exclusion principle which holds for both holons and spinons:
the dispersions of both types of excitations become
incomplete upon doping, i.e., the spinon Fermi surface shrinks
as if the spinons were spin $1/2$ particles,
while simultaneously the top of the holon band is `sawed off'
as if the holons were spinless Fermions.
It should be noted that this is quite naturally
to be expected in that the rapidities for the different
`particles' in the Bethe ansatz
solution both obey a Pauli-like exclusion principle\cite{BaresBlatter}.
This has negative implications for, e.g., slave boson
mean-field calculations, which necessarily have to treat one type
of excitation as a Boson. While spin-charge separation is often
quoted as justification for the mean-field decoupling,
it is obvious that this approximation must fail to reproduce the
excitation spectrum even qualitatively
in 1D, the only situation where spin-charge
separation is really established.\\
For a more quantitative discussion of the Fermi points, 
we note that the Fermi momentum
for hole concentration $\delta$ is $k_F= \frac{\pi}{2}(1-\delta)$. 
For this momentum the first branch of low energy excitations
reaches $E_F$. For small $\delta$ the second branch of low energy excitations
comes up to $E_F$ at $-3k_F + 2\pi= \frac{\pi}{2}(1+ 3\delta)$.
The two marginal singularities thus enclose a hole pocket
of length $2\pi \delta$ as one would expect for
holes corresponding to spinless Fermions. It is easy to see, that this
hole pocket is nothing but the manifestation of the
holon `Fermi surface' around $k$$=$$\pi$: the lowest charge excitations,
which may be thought of as corresponding to a particle hole excitation
between the two edges of the holon pocket have wave vector
$4k_F = 2\pi - 2\pi \delta$, i.e., the holon pocket
has a diameter of $2\pi/\delta$, precisely the distance between
the two marginal singularities. The spectral function for the
doped case thus follows the same building principle as for the
case of half filling, with the sole difference being that
occupied spinon or holon momenta are no longer available
for the construction of final states. 
The singularities in $n(k)$ may be thought of as
enclosing a hole pocket corresponding to spinless Fermions,
and thus reflect the Fermi surface of the holons.
The two holon pockets are placed such that their inner edges
at $\pm k_F$ enclose the volume corresponding to the 
Fermi sea of spinons of density $(1/2)(1-\delta)$.

\section{Two dimensions, doped case}

We proceed to the doped case in 2D.
Let us note from the very beginning that for very simple
technical reasons the situation is much more unfavorable in this
case. To begin with, due to the higher symmetry of 2D clusters 
the available $\vec{k}$ meshes are much coarser:
for example, amongst the $18$ allowed momenta in the $18$ site
cluster only $6$ $\vec{k}$-points are actually non symmetry-equivalent,
so that the amount of nonredundant information is much smaller than
in 1D. Next, unlike 1D where a unique relationship exists
between hole density and Fermi momentum,
most electron numbers in small 2D clusters
correspond to open-shell configurations
with highly degenerate ground states for noninteracting particles.
In an open-shell situation multiplet effects are guaranteed
to occur, so that it is in general unpredictable
which momenta are occupied and which ones are not
(this holds for a Fermi liquid, but is most probably true also for 
other `effective particles'). Unexpected problems may arise from this.
Bearing this in mind, one therefore may not expect to see a similarly 
detailed and clear picture as in 1D.\\
Then, Figure \ref{fig6} shows the photoemission spectra
for the $18$-site cluster,
with two holes. We first consider the left hand panel,
where energies are measured in units of $J$. Comparing
with Fig.~\ref{fig5}, some similarities are quite obvious:
the excitation energies of the topmost peaks at $(0,0)$, $(2\pi/3,0)$ 
are independent of $t$ (although the spectra for $t/J$$=$$2$ show
a slight deviation) so that we can identify a `band' of states
with energy scale $J$. The situation actually is not
entirely clear, in that the peak at $(\pi/3,\pi/3)$ is so
close in energy to the one at $(2\pi/3,0)$ that it is not possible
to decide if their energy difference scales with
$J$ or $t$. Next, the topmost peaks at $(2\pi/3,2\pi/3)$ and
$(\pi/3,\pi)$ show a systematic progression with $t$, which is
very reminiscent of, e.g., Fig.~\ref{fig1}.
Plotting the same spectra with energy scale $t$ indeed to good
approximation aligns these peaks
(although the peak at $(2\pi/3,2\pi/3)$ still has a slight drift),
i.e., their excitation energy relative to the topmost peak
at $(2\pi/3,0)$ scales with $t$. Moreover one can identify
a number of diffuse `features' at energies
between $-0.5t$ and $-t$, which also are roughly aligned;
these are indicated by the dashed line.
In analogy with 1D, we can thus distinguish different branches of states,
with different energy scales in their excitation energies.
While the coarseness of the $\vec{k}$-meshes introduces some
uncertainty, the data are consistent
with a `$J$-band' dispersing upwards in the interior of the
antiferromagnetic Brillouin zone, and a `$t$-band' dispersing 
downwards in the outer part, i.e., the same situation as seen in 1D.
A major difference is the fact that the 
`features' at higher binding energies are all very diffuse,
at least for $J/t$$\ge$$3$. More significantly,
despite the fact that its energy scale seems to be $t$,
the dispersion of the `shadow band' is much weaker than in 1D.
In other words, the effective mass of that band is $\sim$$t^{-1}$,
but with a very large prefactor.\\
We proceed to the $20$-site cluster, 
also doped with two holes (see Fig.~\ref{fig7}).
Choosing $J$ as the unit of energy, we see the
already familiar situation: the topmost peaks
for the states at $(2\pi/5,\pi/5)$ and
$(\pi/5,3\pi/5)$ are aligned (although $t/J$$=$$2$ again
deviates slightly) and several other peaks
show a systematic progression with $t$
(an unexpected exception is $(0,0)$ where a well defined peak
actually is not observed).
Changing to energy scale $t$ aligns
a number of these peaks, which suggests that these peaks
form a `$t$-band' which originates from
the topmost peak at $(2\pi/5,\pi/5)$.
This is a second unexpected feature of the $20$-site cluster, in that for 
the spectra in 1D (and for those of the $18$-site cluster in 2D)
the most intense $t$-band always seemed to originate
from the topmost peak of the photoemission spectrum.
We can only speculate that these unusual features are the consequence of,
e.g., the multiplet effects mentioned above.
We also note in this context that the spectra at $(0,0)$ look actually quite
different for $18$ and $20$ site cluster, which  shows the impact
of finite-size effects.\\
Ascribing the special behavior at $(0,0)$ to finite size-effects, we have
a quite similar picture as in the $18$-site cluster, i.e.,
the topmost peaks for spectra inside the antiferromagnetic zone
have $J$ as their energy scale whereas the topmost peaks
in the outer part of the zone have energy scale $t$
(this also holds for $(\pi,0)$ which is on the boundary
of the antiferromagnetic zone).
As was the case in the $18$ site cluster the `shadow band',
while having $t$ as its energy scale,
has a much weaker dispersion than in 1D.
Indeed, fitting the $t$-bands in both $18$ and $20$ site cluster
by an expression of the form $2t_{eff}(\cos(k_x)+\cos(k_y))$
requires to choose $t_{eff} \approx 0.1t$ - it is tempting to
speculate that this may actually be $\delta t$, as one would expect
e.g. in the Gutzwiller picture. Another notable feature is that the
$t$-band is restricted to the outer part of the
Brillouin zone. Only the diffuse high-energy `band' indicated by the
dashed line in Figure \ref{fig7} seems to scale with $t$.\\
For completeness we would like to mention that a similar analysis was 
not possible
for the $16$-site cluster with $2$ holes. The reason is essentially that
for some momenta there are no more sharp `peaks', but rather
a multitude of densely spaced small peaks. Due to this,
we were not able to assign any  defined `bands', or groups
of peaks which showed a systematic scaling of their excitation energy.
We have also performed this kind of analysis for the $16$ site cluster
with $4$ holes and found no more indication of the energy scale
$J$: at this somewhat higher concentration the entire spectra scale
with $t$.\\
Summarizing the data for 2D, hole doping seems to lead to
behavior which is more reminiscent of 1D than for half filling,
in that the $J$-band dispersing upwards in the inner part of the
Brillouin zone and the $t$-band dispersing downwards in the
outer part seem to exist also in this case. Much unlike 1D,
however, the shadow band, while in principle having $t$ as its energy scale,
still has a very weak dispersion, so that the band structure
in the doped case is practically identical to that
in the undoped system\cite{EderOhtaShimozato}.
We note however, that the fact that the shadow band 
has $t$ as energy scale has profound implications for its explanation:
there have been attempts to interpret the shadow band in Bi2212 as 
a `dynamical replica' of the main band, created by 
scattering of quasiparticles in the standard tight-binding band from
antiferromagnetic spin fluctuations\cite{Chubukov}.
Experimentally, however, the fact that the shadow bands are observed also
in the overdoped compounds\cite{LaRosa}, where antiferromagnetism is
very weak, as well as the fact that they do not seem to become
more pronounced in the underdoped compounds\cite{Marshall}, 
where antiferromagnetism is strong, both suggest otherwise.
On the theoretical side, we believe that our data very clearly
rule out this interpretation: both the `main band'
and the spin correlation function\cite{EderOhtaMaekawa} 
have $J$ as their relevant energy scale,
and it would be very hard to understand how the energy scale of $t$
for the shadow band should emerge from a combination of these two
types of excitations. 
In fact, the relatively accurate scaling with physically
very different parameters suggests completely
different propagation mechanisms for the two types
of excitations. We therefore believe that the shadow band is a
separate branch of excitations, probably best comparable to the
states which produce the $3k_F$ singularity in the 1D systems.

\section{Discussion}

In the previous sections we have investigated the photoemission spectrum
for the one and two dimensional $t$$-$$J$ model. By studying the parameter 
dependence of the spectra, we could in 1D identify
`branches' of states which trace out the dispersions of the
elementary excitations of the TLL, the spinons and holons.
Both elementary excitations have a simple nearest neighbor hopping
dispersion, but with different band width: that of the spinons
is $\sim$$J$, that of the holons $\sim$$t$.\\
In the doped case there are two groups of
states which touch the Fermi energy (see Figure \ref{figx}).
`Inside' the noninteracting Fermi surface, there is a 
whole continuum of bands dispersing upwards to $E_F$.
The uppermost of these bands 
traces out the spinon dispersion and has $J$ as its energy scale,
the lowermost band traces out the holon dispersion and has $t$ as its
energy scale. In the thermodynamic limit these bands degenerate
into `cusps' and merge at $E_F$.
In the outer half of the Brillouin zone, there are only
states which have $t$ as their energy scale.
These reach the Fermi energy at $3k_F$,
giving rise to a second Fermi point. While the resolution
in $k$ and $\omega$ available in our finite clusters is not sufficient
to make statements about extreme low energy excitations,
the positions of the singularities in the
electron momentum distribution as determined from exact solutions
clearly shows that both branches of states indeed do touch $E_F$.
The two singularities may be thought of enclosing a hole pocket
of extent $2\pi\delta$, which is essentially the image of the
holon Fermi surface.\\
In 2D at half-filling, the situation is quite different:
while it is still possible to distinguish bands
with different scaling behavior with $J$ and $t$,
the spinon Fermi surface present in 1D does not exist
and the `holons' seem to correspond to overdamped resonances
rather than sharp excitations as in 1D. We propose that
strong attraction between spin and charge excitations,
most probably due to the well-known string mechanism,
pulls a band of bound states out of the continuum of
`free' holon and spinon states. The relevant physics thus is that of
spin-bag--like spin 1/2 quasiparticles, as suggested by
a considerable amount of numerical evidence.\\
For the doped case in 2D, the situation is less clear
and actually somewhat ambiguous. The numerical photoemission spectra
show some analogy with 1D, in that there seems to be
a high intensity `main band'  with energy scale $J$
dispersing upwards in the inner part of the Brillouin zone, and a
low intensity `shadow band' with energy scale
$t$ dispersing downwards in the outer part of the Brillouin zone
(see Figure \ref{figx}).
In contrast to 1D the dispersion of the
shadow band is much weaker, i.e., while the energy scale of the dispersion
is $t$, it has an additional very small prefactor (of the order of the
hole concentration).
Moreover the $t$-band seems limited to the outer part of the
Brillouin zone, i.e. there are no indications for a holon band with energy 
scale $t$ dispersing upwards in the inner part of the Brillouin zone.
Only in the $18$-site cluster a diffuse `band' with energy scale
$t$ can be roughly identified at higher binding energies.
The different energy scales
of main band and shadow band suggest that these are excitations
of quite different nature, and in particular rule out
the explanation that that the shadow band is
created by scattering from antiferromagnetic spin fluctuations.\\
Turning to experiment, the results for 2D immediately
suggests a comparison
with the data of Aebi { \em et al.}\cite{Aebi}. These authors found that
in addition to the `bright' part of the band structure, which seems
to be consistent with the noninteracting one,
there is also a low intensity `replica', shifted approximately
by $(\pi,\pi)$,
which had been consistently overlooked in all previous studies.
If one wants to make a correspondence to the situation for the
$t$$-$$J$ model, one thus should identify this low intensity
part with the $t$-band dispersing downwards in the $t$$-$$J$ model.
Our data imply that the shadow band should have a slightly different
dispersion than the main band. The limitations of the cluster
method probably preclude any meaningful quantitative statements,
but it might be interesting to see if this difference in dispersion
can be resolved experimentally.\\
We conclude by outlining a somewhat speculative scenario, based on the
assumption that the two bands represent indeed different excitations,
which persist at all temperatures and independent of antiferromagnetic
correlations. In this case, the topology in 2D opens an 
interesting possibility:
whereas in 1D the two classes of low energy excitations 
forming the $k_F$ and $3k_F$ singularities in $n(\vec{k})$ are
well separated in $\vec{k}$ and $\omega$ space for simple topological
reasons, the
experimental data of Aebi {\em et al.} indicate that the
main and shadow band intersect at certain points in the 
Brillouin zone (see Fig.~\ref{fig9}, left panel).
Neglecting the small difference in dispersion between main and
shadow band, we might therefore model the low energy excitation spectrum by
the effective Hamiltonian
\begin{equation}
H_{QP} = \sum_{\vec{k},\sigma}
(\;\epsilon(\vec{k})\;
a_{\vec{k},\sigma}^\dagger a_{\vec{k},\sigma} +
\epsilon(\vec{k}+\vec{Q})\;
b_{\vec{k},\sigma}^\dagger b_{\vec{k},\sigma}\;)
\label{eff}
\end{equation}
where $\epsilon(\vec{k})$ is the dispersion of the main band, $\vec{Q}
=(\pi,\pi)$, and the $a$ and $b$ operators refer to the main and shadow band,
respectively. 
Choosing a dispersion of the form
$\epsilon(\vec{k})= -2t(\cos(k_x) + \cos(k_y)) + 
4t'\cos(k_x)\cos(k_y)$ this Hamiltonian reproduces the Fermi surface
topology found by Aebi {\em et al.} quite well (see Fig.~\ref{fig9}). 
However, as mentioned above
the two branches of excitations intersect at some points of the Brillouin zone,
so that already a small mixing between the two bands, which in turn
may originate from the spinon-holon interaction, has a dramatic
effect on the topology of the
low energy excitation spectrum. Namely
adding a term of the form
\[
H_{mix}= \Delta \sum_{\vec{k},\sigma} 
(\; a_{\vec{k},\sigma}^\dagger b_{\vec{k},\sigma} + {\rm H.c.}\;),
\]
i.e., a hybridization between the two types of bands, even
relatively small values of $\Delta$ open up a gap around $(\pi,0)$
and transform the Fermi surface transformed into a hole
pocket (see left panel of Fig.~\ref{fig9}).
Thereby we fix the chemical potential by requiring that the number of
$a$ and $b$ particles remains unchanged; it is easy to see that the
area covered by the pockets then equals the hole concentration
$\delta$, precisely as it was the case in 1D.
Thereby the Fermi surface has predominant main band character
at its inner edge, and shadow band character at the outer edge,
implying a very different `visibility' in photoemission.
Finally, it is tempting to speculate that the
`pseudo gap order parameter' $\Delta$ decreases
with increasing temperature/hole concentration.
Its vanishing at a certain temperature $T^*$ then could
produce a crossover from the hole pockets to the `large' Fermi surface
at $T^*$, which picture would nicely reproduce the
pseudogap phenomenology observed\cite{Loeser} in cuprate superconductors.\\
Financial support of R. E. by the European Community and of Y. O. by the
Saneyoshi Foundation and a Grant-in-Aid for Scientific Reserach from the
Ministry of Education, Science and Culture of Japan is
most gratefully acknowledged.
%%%%%%%%%%%%%%%%%%%%%%%%%%%%%%%%%%%%%%%%%%%%%%%%%%%%%%%%%%%%%
\begin{figure}
\narrowtext
\caption[]{
Photoemission spectra at half-filling for different values
of $J/t$. The thin lines mark the spinon and holon band.}
\label{fig1}
\end{figure}
%%%%%%%%%%%%%%%%%%%%%%%%%%%%%%%%%%%%%%%%%%%%%%%%%%%%%%%%%%%%%
%%%%%%%%%%%%%%%%%%%%%%%%%%%%%%%%%%%%%%%%%%%%%%%%%%%%%%%%%%%%%
\begin{figure}
\narrowtext
\caption[]{Photoemission spectra at half-filling
in the $12$-site ring. For all $J/t$ the energy of the state marked by
the black dot is taken as the zero of energy.}
\label{fig2}
\end{figure}
%%%%%%%%%%%%%%%%%%%%%%%%%%%%%%%%%%%%%%%%%%%%%%%%%%%%%%%%%%%%%
%%%%%%%%%%%%%%%%%%%%%%%%%%%%%%%%%%%%%%%%%%%%%%%%%%%%%%%%%%%%%
\begin{figure}
\narrowtext
\caption[]{(a) Schematic building principle of the spectral function
at half-filling: Spinon band (thick dashed lines), and holon bands
(thin full lines).
(b) Theoretical spectrum overlaid with the numerical result
for the $20$-site ring with $J/t$$=$$10$. 
The centers of the circles give the
excitation energies and their diameter the pole strength of the peaks
in the electron removal spectrum.}
\label{fig3}
\end{figure}
%%%%%%%%%%%%%%%%%%%%%%%%%%%%%%%%%%%%%%%%%%%%%%%%%%%%%%%%%%%%%
%%%%%%%%%%%%%%%%%%%%%%%%%%%%%%%%%%%%%%%%%%%%%%%%%%%%%%%%%%%%%
\begin{figure}
\narrowtext
\caption[]{Allowed momenta for the $20$-site cluster.}
\label{fig3a}
\end{figure}
%%%%%%%%%%%%%%%%%%%%%%%%%%%%%%%%%%%%%%%%%%%%%%%%%%%%%%%%%%%%%
%%%%%%%%%%%%%%%%%%%%%%%%%%%%%%%%%%%%%%%%%%%%%%%%%%%%%%%%%%%%%
\begin{figure}
\narrowtext
\caption[]{Photoemission spectra at half-filling for the
2D $20$-site cluster. For all $J/t$ the zero of energy has been set to
the energy of the peak marked by the black dot.}
\label{fig4}
\end{figure}
%%%%%%%%%%%%%%%%%%%%%%%%%%%%%%%%%%%%%%%%%%%%%%%%%%%%%%%%%%%%%
%%%%%%%%%%%%%%%%%%%%%%%%%%%%%%%%%%%%%%%%%%%%%%%%%%%%%%%%%%%%%
\begin{figure}
\caption[]{Photoemission spectra for the $12$-site ring with $2$ holes.}
\label{fig5}
\end{figure}
%%%%%%%%%%%%%%%%%%%%%%%%%%%%%%%%%%%%%%%%%%%%%%%%%%%%%%%%%%%%%
%%%%%%%%%%%%%%%%%%%%%%%%%%%%%%%%%%%%%%%%%%%%%%%%%%%%%%%%%%%%%
\begin{figure}
\caption[]{Photoemission spectra for the 
2D $18$-site cluster with $2$ holes. 
The spectra for momenta
marked by an asterisk have been multiplied by a factor of $2$ for
clarity.}
\label{fig6}
\end{figure}
%%%%%%%%%%%%%%%%%%%%%%%%%%%%%%%%%%%%%%%%%%%%%%%%%%%%%%%%%%%%%
%%%%%%%%%%%%%%%%%%%%%%%%%%%%%%%%%%%%%%%%%%%%%%%%%%%%%%%%%%%%%
\begin{figure}
\narrowtext
\caption[]{
Photoemission spectra for the 2D $20$-site cluster with $2$ holes.
The spectra for momenta marked by an asterisk have been multiplied by a 
factor of $2$ for clarity.}
\label{fig7}
\end{figure}
%%%%%%%%%%%%%%%%%%%%%%%%%%%%%%%%%%%%%%%%%%%%%%%%%%%%%%%%%%%%%
%%%%%%%%%%%%%%%%%%%%%%%%%%%%%%%%%%%%%%%%%%%%%%%%%%%%%%%%%%%%%
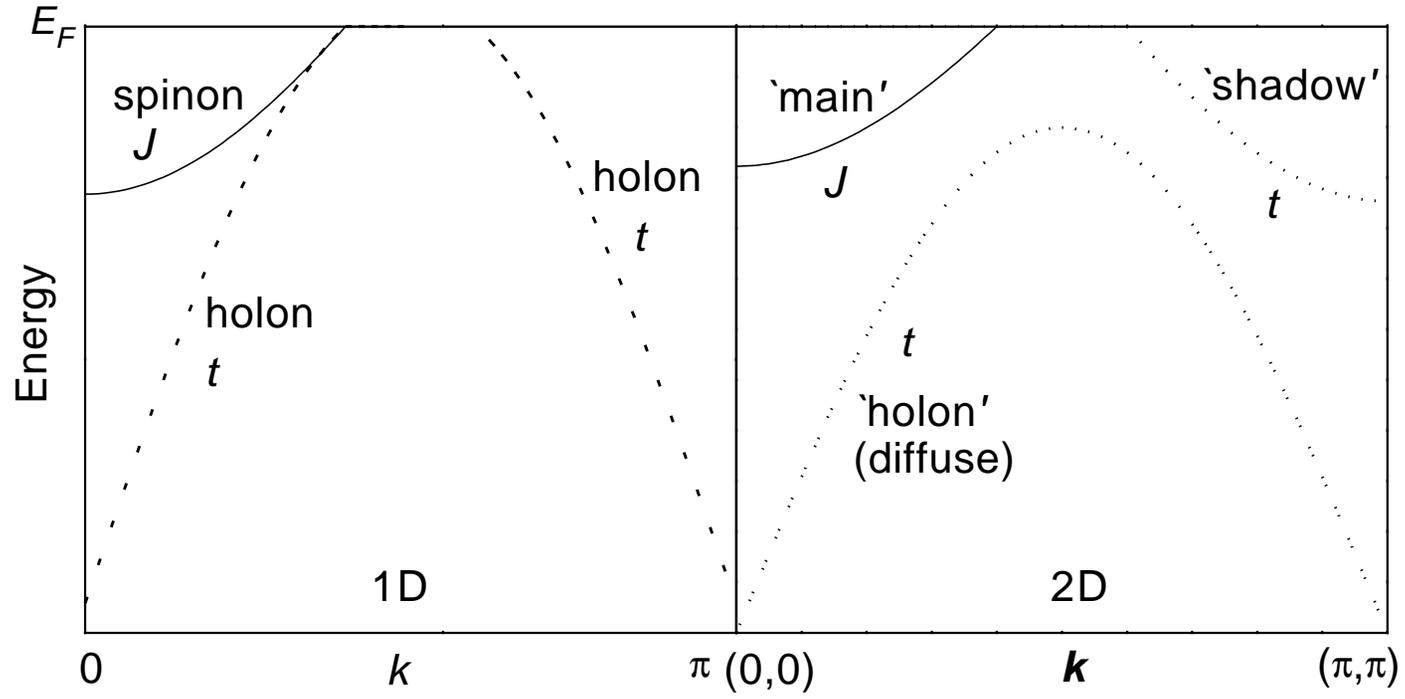
\begin{figure}
\narrowtext
\caption[]{Schematic representation of the `band structure'
in 1D and 2D systems as deduced from the diagonalization spectra.}
\label{figx}
\end{figure}
%%%%%%%%%%%%%%%%%%%%%%%%%%%%%%%%%%%%%%%%%%%%%%%%%%%%%%%%%%%%%
%%%%%%%%%%%%%%%%%%%%%%%%%%%%%%%%%%%%%%%%%%%%%%%%%%%%%%%%%%%%%
\begin{figure}
\narrowtext
\caption[]{
Left hand panel: Fermi surface for the Hamiltonian (\ref{eff})
with $t'/t=0.3$, the full line corresponds to
the main band, the dashed line to the shadow band.
Right hand panel: Fermi surface for $t'/t$$=$$0.3$, $\Delta/t$$=$$0.2$.}
\label{fig9}
\end{figure}
%%%%%%%%%%%%%%%%%%%%%%%%%%%%%%%%%%%%%%%%%%%%%%%%%%%%%%%%%%%%%

\end{multicols}
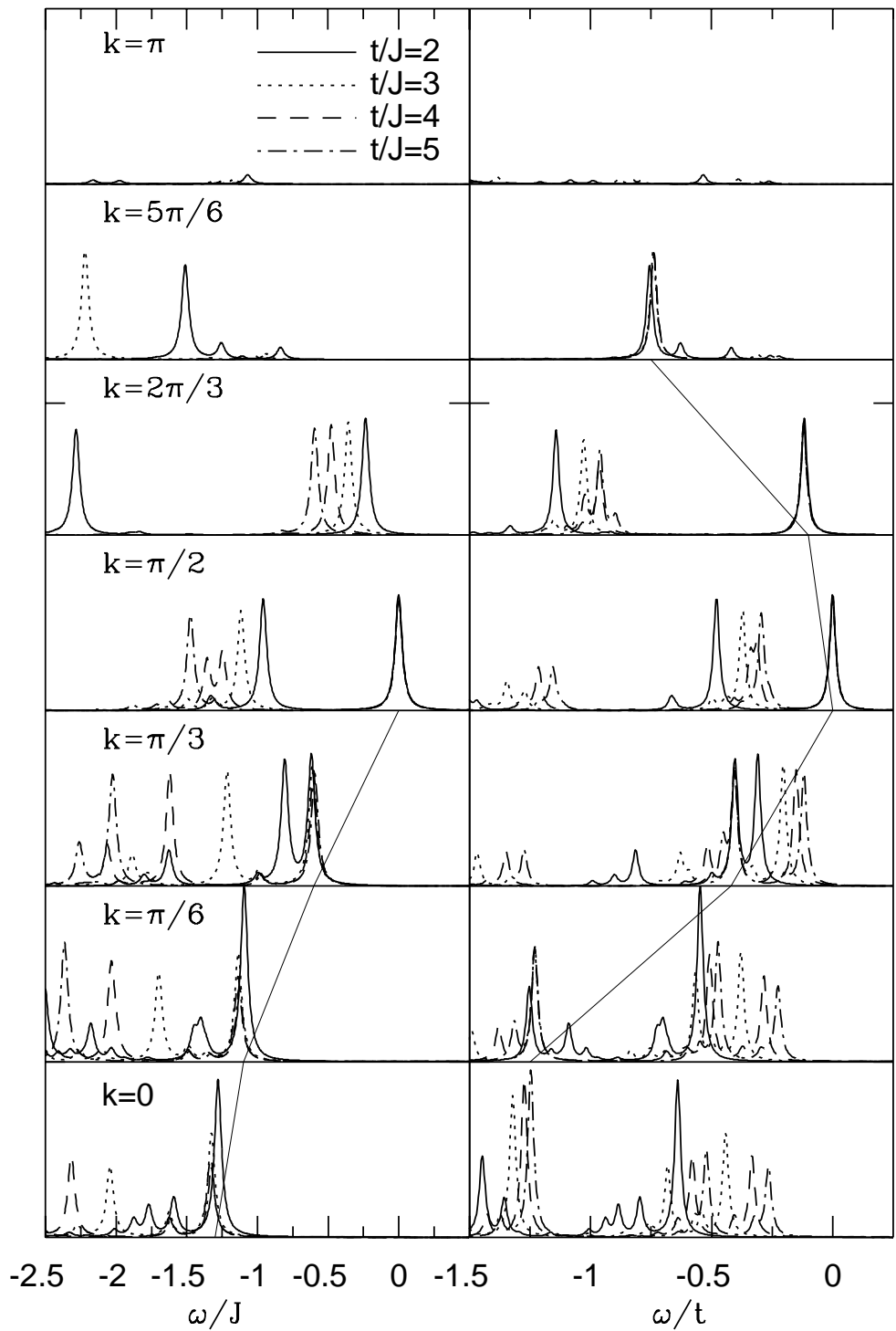
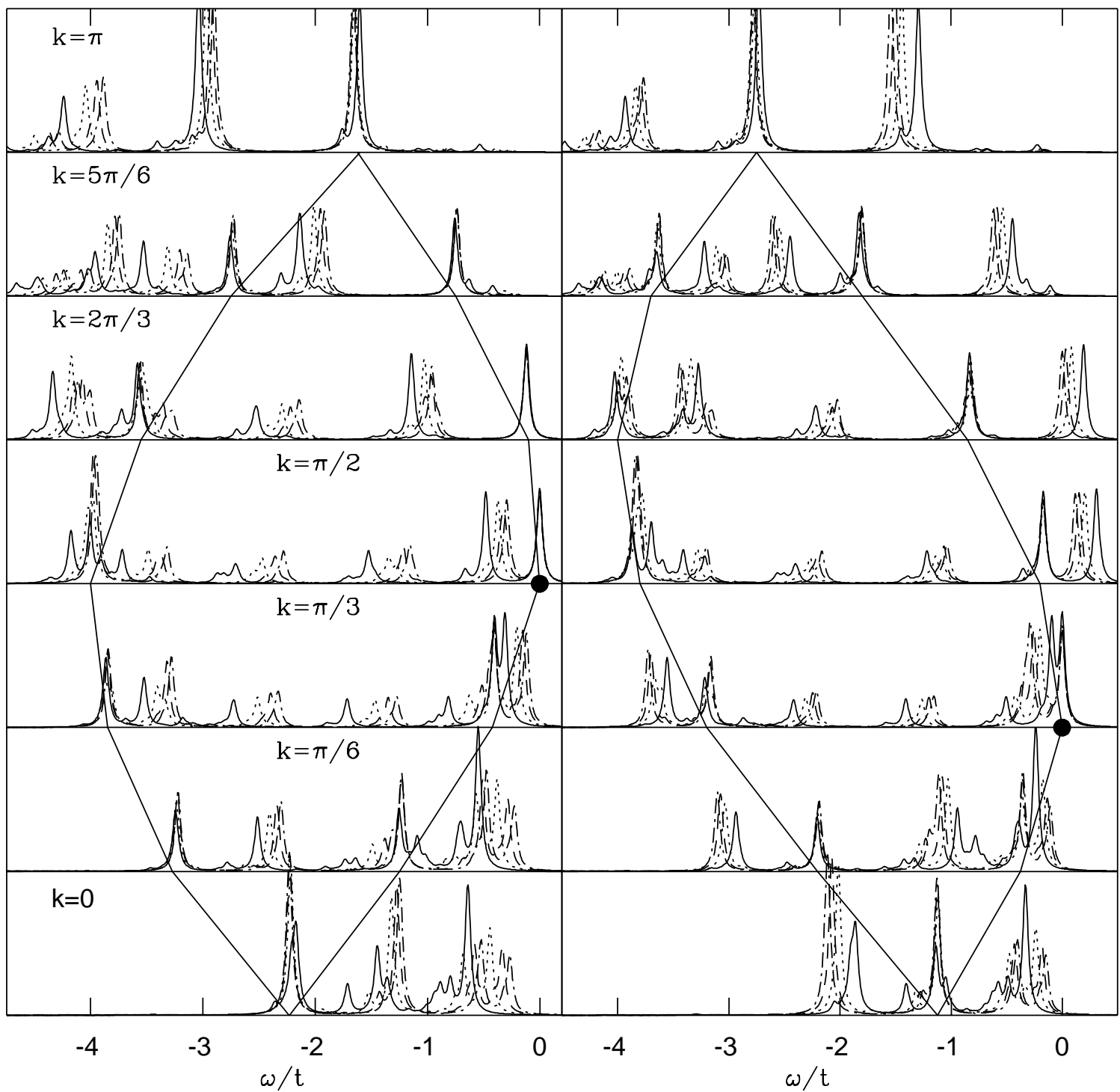
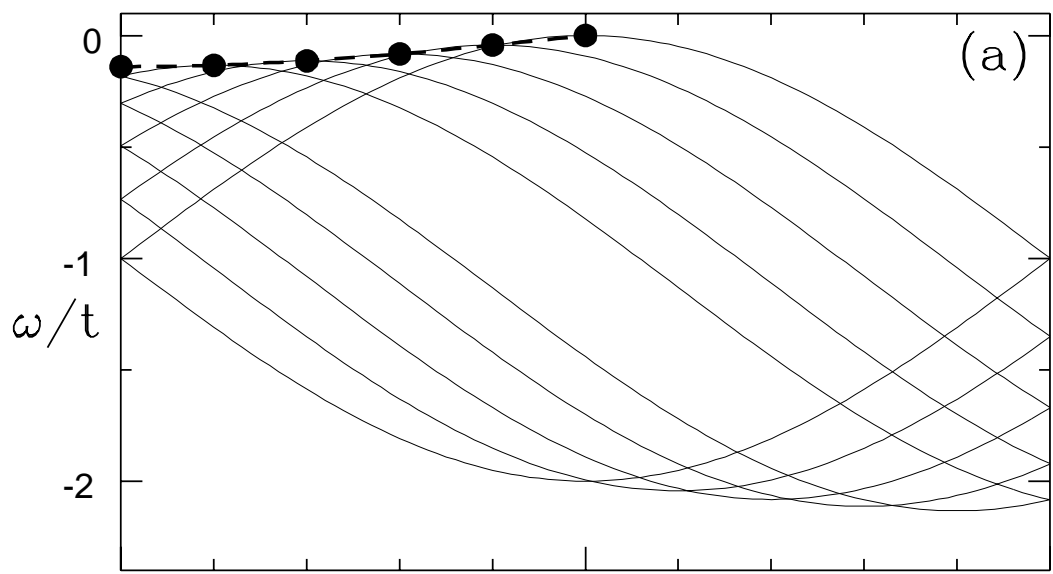
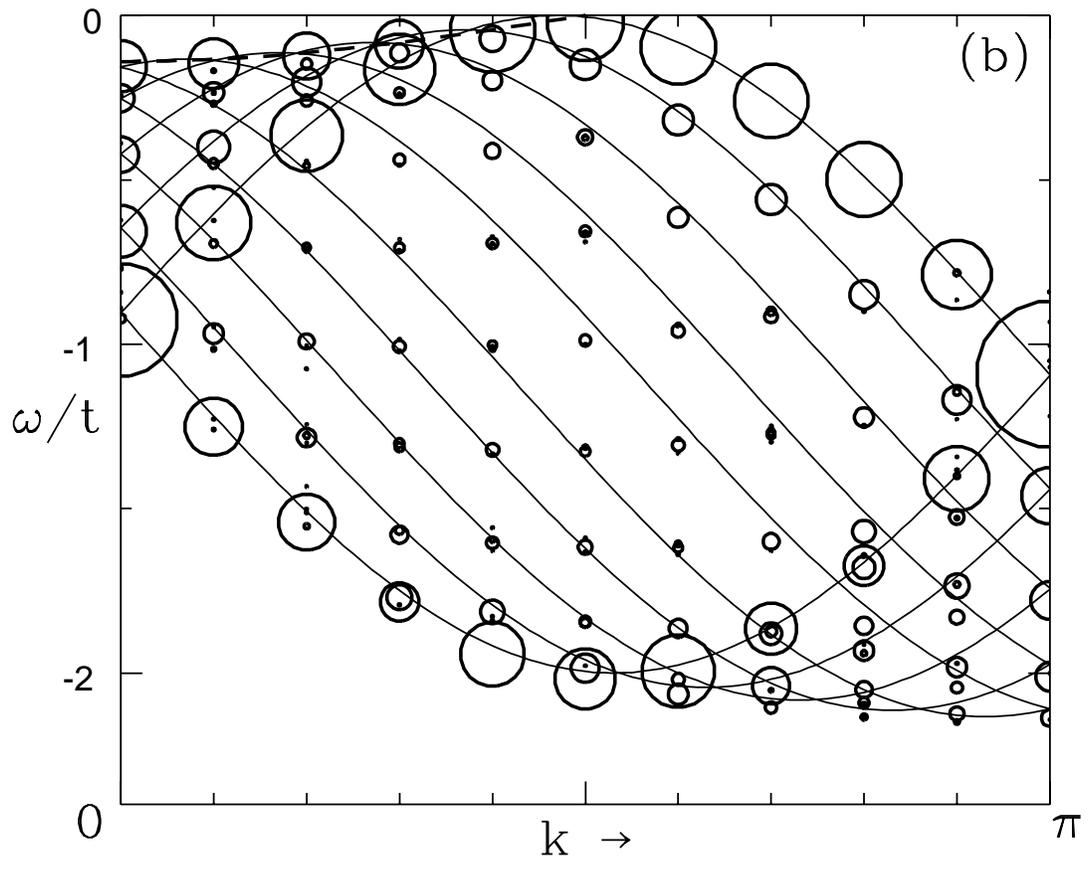
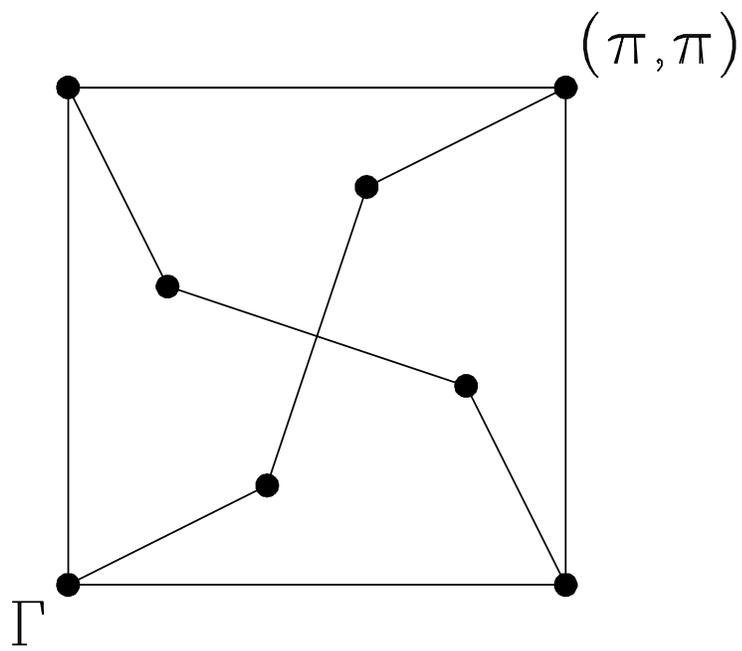
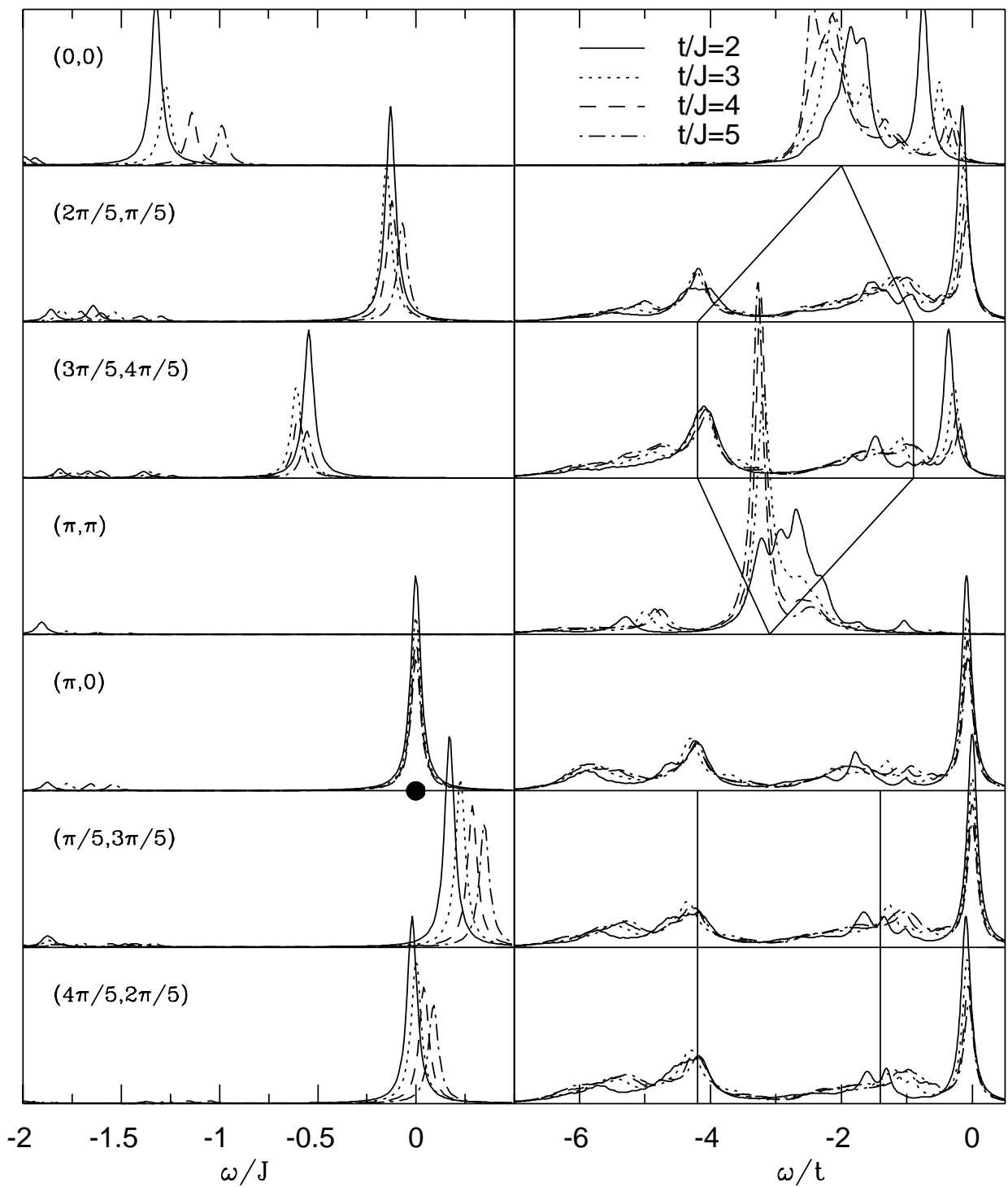
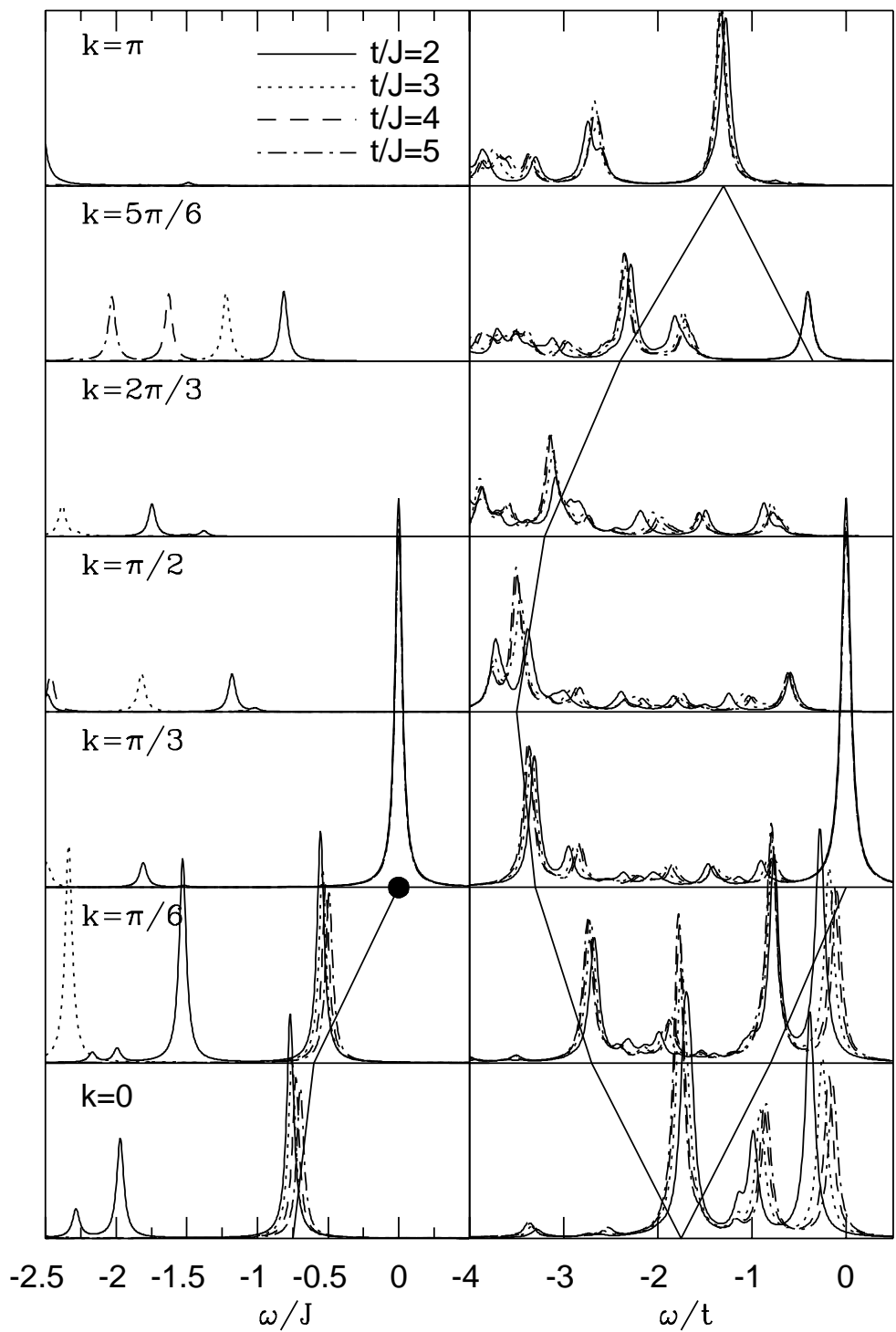
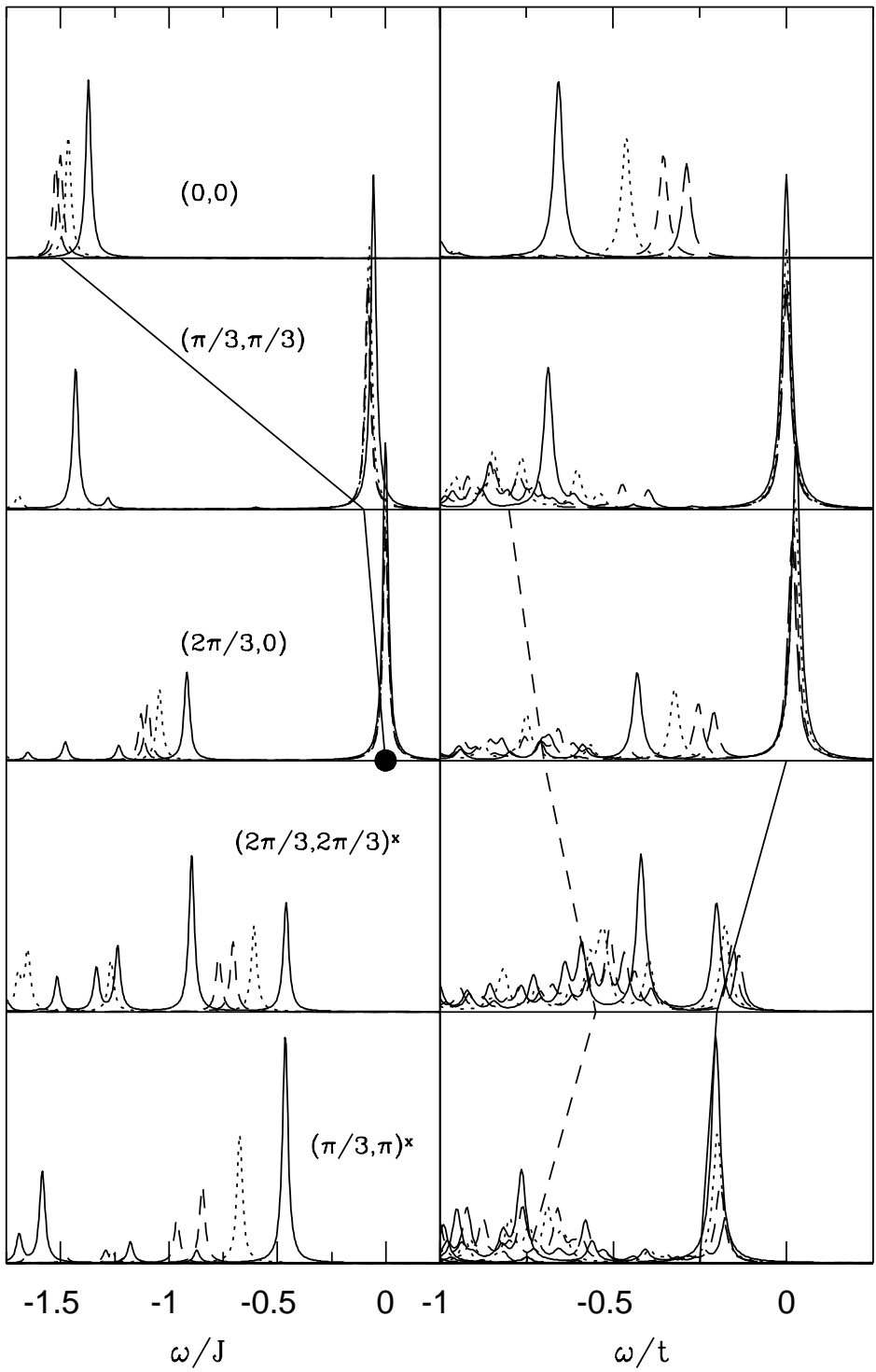
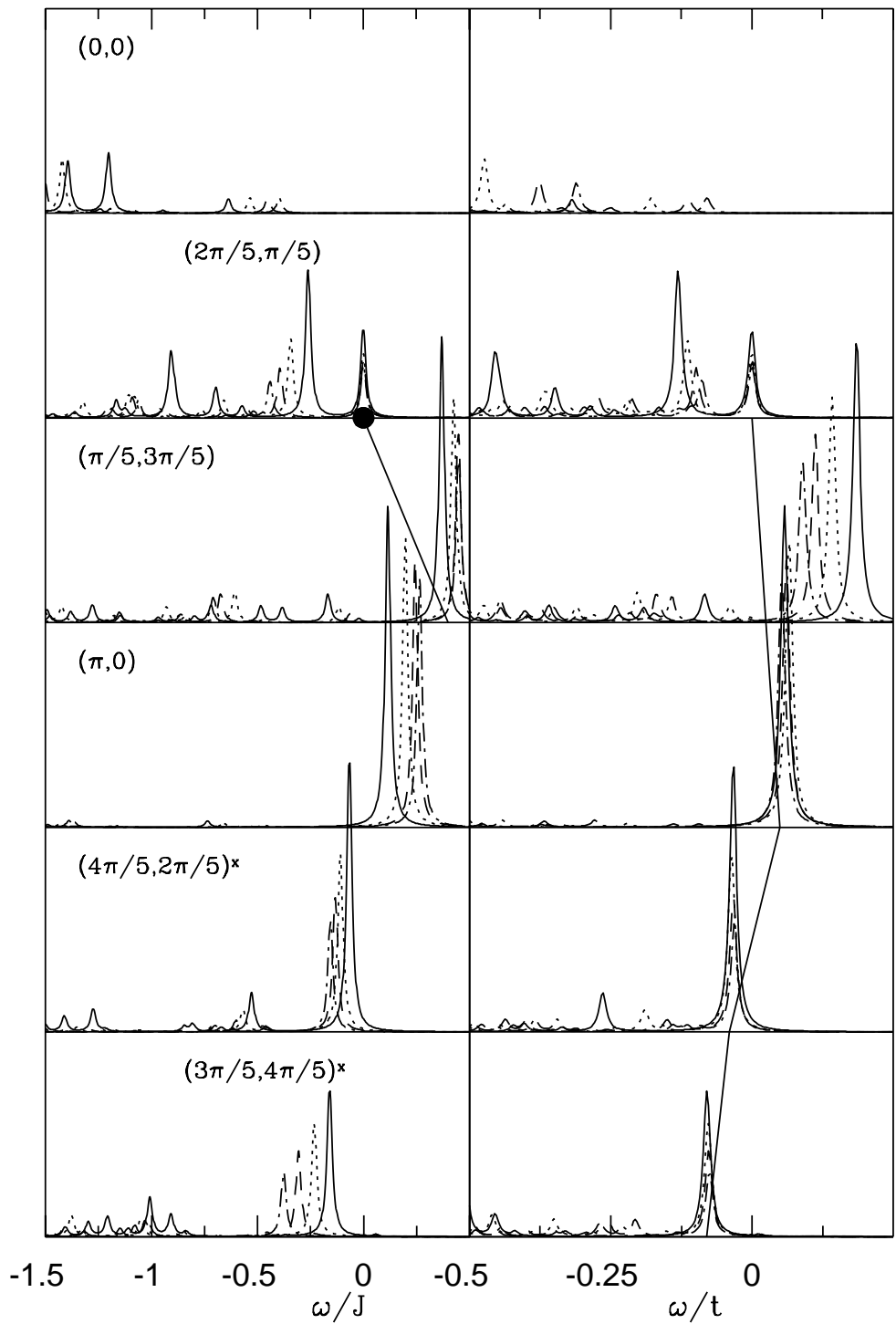
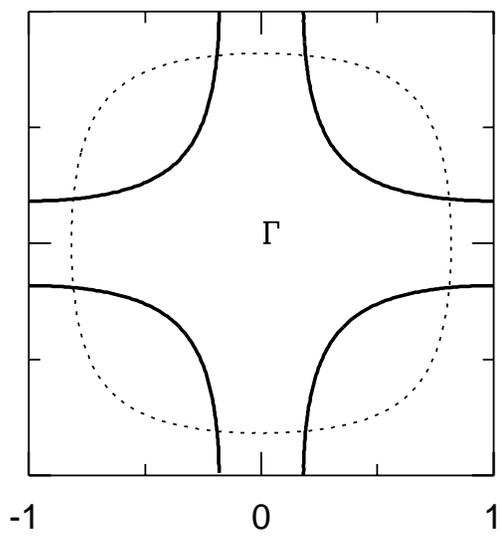
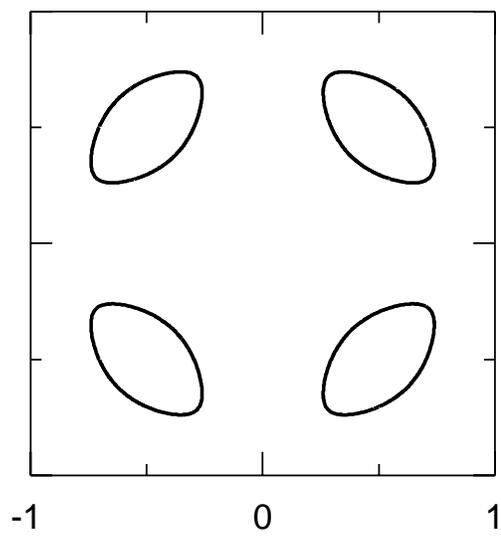
\end{document}